\documentclass[twocolumn,preprintnumbers,superscriptaddress]{revtex4}
\usepackage{amsmath}
\usepackage{amsfonts}
\usepackage{dcolumn}
\usepackage{bm}
\usepackage{epsfig}
\usepackage{epstopdf}
\usepackage{graphicx}
\usepackage{amsfonts}
\usepackage{amssymb}
\usepackage{mathrsfs}
\usepackage{color}
\usepackage{appendix}
\usepackage{multirow}

\begin{document}

\textbf{Xiong {\it et al} Reply:} Recently we reported an experimental verification of an information-theoretic equality \cite{prl-120-010601},
\begin{equation}
\langle e^{-I_{nm}} \rangle :=\sum_{nm} p_{nm} e^{-I_{nm}} =1,
\label{Eq1}
\end{equation}
which is relevant to the Jarzynski equality $\langle e^{\beta(W-\Delta F)}\rangle=1$,
as predicted by a previous theory \cite{vedral,vedral1}.

The Comment \cite{axive1} by Campisi and H$\ddot{\text{a}}$nggi argues that an equation of the mutual information $I_{nm}=-\beta(W-\Delta F)$, which is Eq. (4) in \cite{prl-120-010601} as the connection between Eqs. (1) and (2), is not generally valid, but only under specific conditions.
We consider that the comment involves something misleading, and clarifying some important points is necessary.

\begin{figure}[htpb]
\centering {\includegraphics[width=8.5 cm, height=2.5 cm]{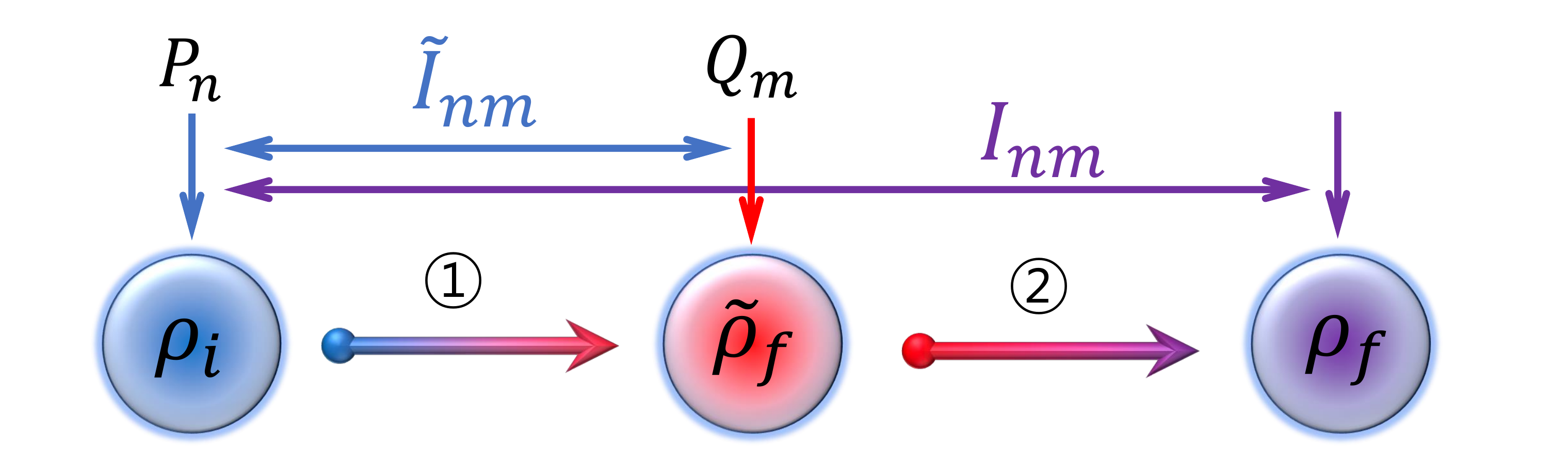}}
\caption{Schematics of two steps for reaching $I_{nm}=-\beta(W-\Delta F)$ from the initial state $\rho_i$. $\tilde{I}_{nm}$ is what
the Comment \cite{axive1} presents due to lack of a thermalization step $\textcircled{2}$ to reach $\rho_f$.}
\label{Fig1}
\end{figure}

First, the final result, i.e., Eq. (8), in \cite{axive1} is a straightforward result of our theory, for which we had already obtained when writing Refs. \cite{prl-120-010601,vedral,vedral1}. Simply speaking, using Eq. (2) in \cite{prl-120-010601}, where $p_{m|n}=\text{Tr}[Q_mP_n]$ and $q_m=\sum_ne^{-\beta E_n^i} \text{Tr}[Q_mP_n]/\sum_ne^{-\beta E^i_n}$, we can easily obtain
\begin{equation}
\tilde{I}_{nm}=\ln \frac{(\sum_ne^{-\beta E^i_n})\text{Tr}[Q_mP_n]}{\sum_n e^{-\beta E_n^i} \text{Tr}[Q_mP_n]},
\label{Eq3}
\end{equation}
which is actually Eq. (8) in \cite{axive1} but contains no interesting physics. Note that our purpose in Refs. \cite{prl-120-010601,vedral,vedral1} is to bridge a relationship between quantum information and thermodynamics. To this end, we need to go a further step after reaching the above equation, as plotted in Fig. \ref{Fig1}. This further step is a thermalization of the state to the canonical distribution as mentioned in \cite{vedral1}. Here we explain it again in a more clarified way. With Gibbs state prepared in \cite{prl-120-010601}, after projecting the initial thermal state onto $P_n$ followed with a free evolution, we quench with $H_f$, and the state finally thermalizes to the canonical distribution $\rho_f=e^{-\beta H_f}/Z_f$  with $Z_f=\sum_n e^{-\beta E^f_n}$, implying $\sum_{nm} p_{m|n} e^{-\beta E_m^f}/Z_f=1$. Thus we have
\begin{eqnarray}
&&\sum_{nm}p_{m|n}\frac{e^{-\beta E_m^f}}{Z_f}=\sum_{nm}p_{nm}e^{-\ln p_n}\frac{e^{-\beta E_m^f}}{Z_f} \notag \\
&&\qquad =\sum_{nm}p_{nm}e^{-\beta (E_m^f-E_n^i)-\beta\Delta F}=1,
\end{eqnarray}
which is the Jarzynski equality \cite{prl-78-2690} and relevant to Eq. (\ref{Eq1}) provided that the mutual information is written as $I_{nm}=-\beta(W-\Delta F)$.

Second, we have to say that the Comment \cite{axive1} introduces 'open quantum system' into our theory, which is unnecessary. In fact, no decoherence is considered in both our theory and experiment \cite{explain}. To avoid decoherence, the thermalization in a quantum system is to reach an equilibrium state, which is accomplished in a closed system. Specifically, in \cite{prl-120-010601}, after free evolution as step $\textcircled{1}$ in Fig. 1, we obtained a non-equilibrium state of $H_{f}$. Thermalizing the state to equilibrium is just the transformation to the equilibrium state $\rho_{f}$ of $H_{f}$ by unitary operations. In addition, we worked with two-level Gibbs states, which means $\Delta F=0$. As such, the work between the initial and final states could be simply obtained by the difference of the corresponding eigenvalues as $W=E_n^i-E^f_m$.

In summary, the confusion made in Comment \cite{axive1} includes omission of the state thermalization to equilibrium and misunderstanding of our work carried out in an open quantum system. Here we would like to emphasize again that Eq. (\ref{Eq1}) can be relevant to the Jarzynski equality under the condition that the initially prepared Gibbs states, under measurements with $P$ and $Q$, thermalize to a canonical distribution in the absence of decoherence. Experimental work in \cite{prl-120-010601}, based on the simplest Gibbs states in a qubit, shows the possibility to bridge the information-theoretic equality to the Jarzynski equality.

\noindent T. P. Xiong$^{1}$, L. L. Yan$^{1}$, F. Zhou$^{1}$, K. Rehan$^{1}$, D. F. Liang$^{1}$, L. Chen$^{1}$, W. L. Yang$^{1}$, Z. H. Ma$^{2}$, M. Feng$^{1}$, and V. Vedral$^{3}$ \\
$^{1}$ Wuhan Institute of Physics and Mathematics, Chinese Academy of Sciences, Wuhan, 430071, China \\
$^{2}$ Department of Mathematics, Shanghai Jiaotong University, Shanghai, 200240, China \\
$^{3}$ Department of Physics, Clarendon Laboratory, University of Oxford, Parks Road, Oxford OX1 3PU, United Kingdom


\begin{thebibliography}{99}
\bibitem{explain} In our expriment, dephasing was employed to produce Gibbs states, which is the starting point of our work. Except this initial state preparation, the whole procedure of operations was carried out under the negligible decoherence, since our implementation time (less than 50 $\mu$s) is much shorter than the decoherence (dephasing) time of the qubit [i.e., T$_{1}$(T$_{2})=$ 1 s (2 ms)].
\end{thebibliography}
\end{document}